\documentclass[epsf,showpacs]{revtex4}
\usepackage{epsf}
\usepackage{graphicx}

\begin{document}

\title{Diffusion in supersonic,  turbulent, compressible flows}
 \author{Ralf S.\ Klessen}
 \affiliation{Astrophysikalisches Institut Potsdam, An der
 Sternwarte 16, 14482 Potsdam, Germany}
 \affiliation{UCO/Lick Observatory, University of California at
 Santa Cruz, Santa  Cruz, CA 95064, U.S.A.}
\email{rklessen@aip.de}
 \author{Douglas N.\ C.\ Lin}
 \affiliation{UCO/Lick Observatory, University of California at
 Santa Cruz,\\ Santa
 Cruz, CA 95064, U.S.A.}
\email{lin@ucolick.org}

\begin{abstract}
  We investigate diffusion in supersonic, turbulent, compressible
  flows.  Supersonic turbulence can be characterized as network of
  interacting shocks. We consider flows with different rms Mach
  numbers and where energy necessary to maintain dynamical equilibrium
  is inserted at different spatial scales. We find that turbulent
  transport exhibits super-diffusive behavior due to induced bulk
  motions. In a comoving reference frame, however, diffusion behaves
  normal and can be described by mixing length theory extended into
  the supersonic regime. 
  \hfill{\em (Accepted for publication in  Phys.\ Rev.\ E)}
\end{abstract}


\maketitle

\section{Introduction}
\label{sec:introduction}
Laboratory and terrestrial gases and liquids are usually well
described by incompressible flows (e.g.\ \cite{Lesieur}).  In
contrast, the dynamical behavior of typical astrophysical gases, are
characterized by poorly understood highly compressible supersonic
turbulent motion (see e.g.\ \cite{VS00}).  For example, the large
observed linewidths in large molecular clouds show direct evidence for
the presence of chaotically oriented velocity fields with magnitudes
in excess of the sound speed. This random motion carries enough
kinetic energy to counterbalance and sometimes overcompensate the
effects of self-gravity of these clouds \cite{WBM00}.  The intricate
interplay between supersonic turbulence and self-gravity determines
the overall dynamical evolution of these clouds and their observable
features such as their density structure, the star formation rate
within them, and their lifetimes.  Thus, it is importance for the
description of many astrophysical systems to understand in detail the
momentum and heat transfer properties of compressible turbulent gases.

Some important clues on the nature and efficiency of mixing associated
with the clouds' supersonic turbulence can be constrained by the
observed metallicity distribution of the stars formed within them.  In
the Pleiades cluster, stars which emerged from the same molecular
cloud have nearly identical metal abundance \cite{W02}.  This astronomical
context therefore imposes a strong motivation for a general analysis
of the transport and mixing processes in compressible supersonically
turbulent media. 

Analytical and numerical studies of diffusion processes are typically
restricted to certain families of statistical processes, like random
walk \cite{MK00} or remapping models or certain Hamiltonian systems
\cite{I92}. The direct numerical modeling of turbulent physical flows
mostly concentrates on incompressible media (e.g.\ \cite{DS97, MKM99,
OL01}), but some studies have been extended into the weakly
compressible regime \cite{CKM95, HCB95, PPW92, PPW94, PPSW99}.
Although highly compressible supersonic turbulent flows have been
studied in several specific astrophysical contexts \cite{PW00,
SPWHW00, SOG98, OGS99, OSG01, M99, KHM00, SMH00, HMK01, HZMLN01, BM02,
PVP95, PV98, VPP95, VS92, BVS99, BP99, BCP01, PN99, PZN00, BNP02,
GPPL01}, the diffusion properties of such flows have not been
investigated in detail. 

It is the goal of this paper to analyze transport phenomena in
supersonic compressible turbulent flows and to demonstrate that --
analogous to the incompressible case -- a simple mixing length
description can be found even for strongly supersonic and highly
compressible turbulence. We first briefly recapitulate in Section
\ref{sec:formalism} the Taylor formalism for describing the efficiency
of turbulent diffusion in subsonic flows.  In Section
\ref{sec:num-meth} we describe the numerical method which we use to
integrate the Navier-Stokes equation. In Section \ref{sec:transport}
we report the diffusion coefficient obtained in our numerical models,
and in Section \ref{sec:mixing} we introduce an extension of the well
known mixing length approach to diffusion into the supersonic
compressible regime. Finally, in Section \ref{sec:summary} we
summarize our results.

\section{A statistical description of turbulent diffusion}
\label{sec:formalism}

Transport properties in fluids and gases can be characterized by
studying the time evolution of the second central moment of some
representative fluid-elements' displacement in the medium,
\begin{equation}
\label{eqn:def-sigma}
\xi^2_{\vec{r}}(t-t') =   \langle [\vec{r}_i(t)-\vec{r}_i(t')]^2
\rangle_i \;,
\end{equation}
where the average $\langle \cdot \rangle_i$ is taken over an ensemble
of passively advected tracer particles $i$ (e.g. dye in a fluid, or
smoke in air) that are placed in the medium at a time $t'$ at
positions $\vec{r}_i(t')$; or where the average is taken over the
fluid molecules themselves (or equivalently, over sufficiently small
and distinguishable fluid elements). The dispersion in one spatial
direction, say along the $x$-coordinate, is $\xi^2_{x}(t-t') = \langle
[x_i(t)-x_i(t')]^2\rangle_i$. For isotropic turbulence it follows that
$\xi^2_{x} = \xi^2_{y} = \xi^2_{z} = 1/3 \;\xi^2_{\vec{r}}$.  For
fully-developed stationary turbulence, the initial time $t'$ can be
chosen at random and for simplicity is set to zero in what follows.

The quantity $\xi_{\vec{r}}(t)$ can be associated with the diffusion
coefficient $D$ as derived for the classical diffusion equation,
\begin{equation}
\label{eqn:diff}
\frac{\partial n}{\partial t} = D \vec{\nabla}^2 n\;,
\end{equation}
where $n(\vec{r}_i,t)$ is the probability distribution function (pdf)
for finding a particle $i$ at position $\vec{r}_i(t)$ at time $t$ when
it initially was at a location $\vec{r}_i(0)$.  This holds if the
particle position is a random variable with a Gaussian distribution
\cite{B49}. In the classical sense, $n(\vec{r},t)$ may correspond to
the contaminant density in the medium. Equation \ref{eqn:diff} holds
for normal diffusion processes and for time scales larger than the
typical particles' correlation time scale $\tau$.

In general, however, the Lagrangian diffusion coefficient is time
dependent and can be defined as 
\begin{equation}
\label{eqn:D(t)-1}
D(t) =  \frac{d\xi^2_{\vec{r}}(t)}{dt}= 2 \langle
\vec{r}_i(t) \cdot \vec{v}_i(t) \rangle_i \;,
\end{equation}
where $\vec{v}_i(t)=d\vec{r}_i(t)/dt$ is the Lagrangian velocity of
the particle.  The diffusion coefficient along one spatial direction,
say along the $x$-coordinate, follows accordingly as $D_x =
d\xi^2_{x}(t)/dt = 2 \langle x_i(t) v_{x.i}(t) \rangle_i$. Equation
\ref{eqn:D(t)-1} holds for homogeneous turbulence with zero mean
velocity. From $\vec{r}_i(t) = \vec{r}_i(0) + \int_0^t \vec{v}_i(t')
dt'$ it follows that
\begin{equation}
  \label{eqn:D(t)-2}
  D(t) = 2 \left \langle \left[ \vec{r}_i(0) + \int_0^t \vec{v}_i(t') dt'
  \right] \cdot \vec{v}_i(t)\right \rangle_i
       = 2 \int_0^t \langle
\vec{v}_i(t')\cdot\vec{v}_i(t) \rangle_i dt'
\;.
\end{equation}
The above expression allows us to related $D(t)$ to the trace of the
Lagrangian velocity autocorrelation tensor ${\rm{tr}}\,{\cal C}(t-t') =
\langle \vec{v}_i(t') \cdot \vec{v}_i(t) \rangle_i$ as
\begin{equation}
  \label{eqn:correlation-tensor}
  D(t) = 2 \int_0^t {\rm{tr}}\,{\cal C}(t-t') dt' = 2 \int_0^t {\rm{tr}}\,{\cal C}(t') dt' \;,
\end{equation}
a result which was derived by Taylor already 1921 \cite{T21}. This
formulation has the advantage that it is fully general and that it
allows us to study anomalous diffusion processes. Note, that strictly
speaking any transport process with $\xi_{\vec{r}}(t)$ not growing
linearly in time is called anomalous diffusion. This is always the
case for time intervals shorter than the correlation time $\tau$, but
sometimes anomalous diffusion can also occur for $t\gg\tau$.
If $\xi_{\vec{r}}(t) \propto t^{\alpha}$ and if $\alpha < 1$ transport
processes are called {\em subdiffusive}, if $\alpha >1$ they are
called {\em superdiffusive} \cite{Lesieur, I92, CMMV99, LM00}.
Studying transport processes directly in terms of the particle
displacement, i.e.\ Equation \ref{eqn:def-sigma}, is useful when
attempting to find simple approximations to the diffusion coefficient
$D(t)$ for example in a mixing length approach.

\section{Numerical Method}
\label{sec:num-meth}
In order to utilize the above formalism, we carry out a series of
numerical simulation of supersonic turbulent flows.  A variety of
numerical schemes can be used to describe the time evolution of gases
and fluids. By far the most widely-used and thoroughly-studied class
of methods is based on the finite difference representations of the
equations of hydrodynamics (e.g.\ \cite{P77}). In the most simple
implementation, the fluid properties are calculated on equidistant
spatially fixed grid points in a Cartesian coordinate system. Finite
difference schemes have well defined mathematical convergence
properties, and can be generalized to very complex, time varying,
non-equidistant meshes with arbitrary geometrical properties.
However, it is very difficult to obtain a Lagrangian description,
which is essential when dealing with compressible supersonic
turbulence with a high degree of vorticity.  Methods that do not rely
on any kind of mesh representation at all are therefore highly
desirable.

For the current investigation we use smoothed particle hydrodynamics
(SPH), which is a fully Lagrangian, particle-based method to solve the
equations of hydrodynamics. The fluid is represented by an ensemble of
particles, where flow properties and thermodynamic observables are
obtained as local averages from a kernel smoothing procedure
(typically based on cubic spline functions) \cite{B90,M92}. Each
particle $i$ is characterized by mass $m_i$, velocity $\vec{v}_i$ and
position $\vec{r}_i$ and carries in addition density $\rho_i$,
internal energy $\epsilon_i$ or temperature $T_i$, and pressure $p_i$.
The SPH method is commonly used in the astrophysics community because
it can resolve large density contrasts simply by increasing the
particle concentration in regions where it is needed.  This
versatility is important for handling compressible turbulent flows
where density fluctuations will occur at random places and random
times. The same scheme that allows for high spatial resolution in
high-density regions, however, delivers only limited spatial
resolution in low-density regions. There, the number density of SPH
particles is small and thus the volume necessary to obtain a meaningful
local average tends to be large. Furthermore, SPH requires the
introduction of a von~Neumann Richtmyer artificial viscosity to
prevent interparticle penetration, shock fronts are thus smeared out
over two to three local smoothing lengths. Altogether, the performance
and convergence properties of the method are well understood and
tested against analytic models and other numerical schemes, for
example in the context of turbulent supersonic astrophysical flows
\cite{PRL, KB00, KB01, KHM00}, and its intrinsic diffusivity is
sufficiently low to allow for the current investigation of turbulent
diffusion phenomena \cite{L99}.

To simplify the analysis we assume the medium is infinite and isotropic on
large scales, and consider a cubic volume which is subject to periodic
boundary conditions. The medium is described as an ideal gas with an
isothermal equation of state, i.e.\ pressure $p$ relates to the density $\rho$
as $p=c^2_{\rm s}\rho$ with $c_{\rm s}$ being the speed of sound.  Throughout
this paper we adopt normalized units, where all physical constants (like the
gas constant), total mass $M$, mean density $\langle \rho \rangle$, and the
linear size $L$ of the cube all are set to unity. The speed of sound is
$c_{\rm s}=0.05$, hence, the sound crossing time through the cube follows as
$t_{\rm sound} = 20$.  In all models discussed here, the fluid is represented
by an ensemble of $205\,379$ SPH particles which gives sufficient resolution
for the purpose of the current analysis.

\begin{figure}[th]
\includegraphics[width=15cm]{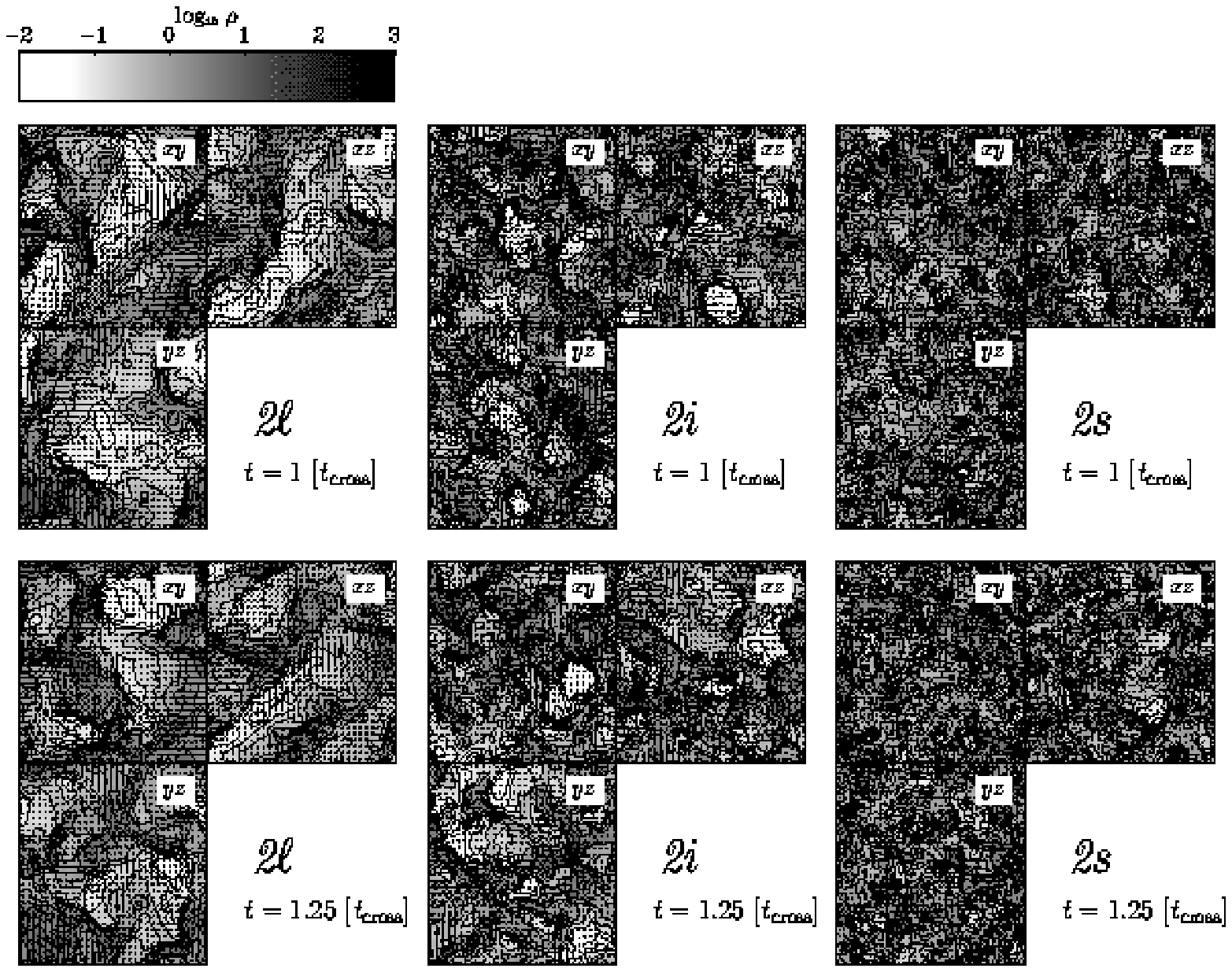}
\caption{\label{fig:3Dplot-cut}%
  Density and velocity structure of models {\em 2}$\ell$, {\em 2i},
  and {\em 2s} (from left to right). The panels show cuts through the
  center of the computational volume normal to the three principal
  axes of the system, after one shock crossing time $t_{\rm cross} =
  L/\sigma_v \approx 6.5$ and 1/4 $t_{\rm cross}$ later. Density is
  scaled logarithmically as indicated in the greyscale key at the
  upper left side. The maximum density is $\sim100$,
  while the mean density is one in the normalized units used.  Vectors
  indicate the velocity field in the plane. The rms Mach number is
  ${\cal M} \approx 3.1$. Large-scale turbulence ({\em 2}$\ell$) is
  dominated by large coherent density and velocity gradients leading a
  large degree of anisotropy, whereas small-scale turbulence ({\em
    2s}) exhibits noticeable structure only on small scales with the
  overall density structure being relatively homogeneous and
  isotropic. }
\end{figure}

Supersonic turbulence is known to decay rapidly \cite{PRL, SOG98, PN99, BM99,
  MB00, BM00}. Stationary turbulence in the interstellar medium therefore
requires a continuous energy input. To generate and maintain the turbulent
flow we introduce random Gaussian forcing fields in a narrow range of
wavenumbers such that the total kinetic energy contained in the system remains
approximately constant. We generate the forcing field for each direction
separately and simply add up the three contributions. Thus, we excite both,
solenoidal as well as compressible modes at the same time. The typical ratio
between the solenoidal and compressible energy component is between 2:1 and
3:1 in the resulting turbulent flow (see e.g.\ Figure 8 in \cite{KHM00}.). We
keep the forcing field fixed in space, but adjust its amplitude in order to
maintain a constant energy input rate into the system compensating for the
energy loss due to dissipation (for further details on the method see
\cite{M99,KHM00}). This non-local driving scheme allows us to exactly control
the (spatial) scale which carries the peak of the turbulent kinetic energy. It
is this property that motivated our choice of random Gaussien fields as
driving source. In reality the forcing of turbulence in the interstellar
medium is likely to be a multi-scale phenomenon with appreciable contributions
from differential rotation (i.e.\ shear) in the Galctic disk and energy input
from supernovae explosions ending the lives of massive stars \cite{MK03}.
Comparable to the values observed in interstellar gas, we study flows with
Mach numbers in the range 0.5 to 10, where we define the Mach number from the
{\em one-dimensional} rms velocity dispersion $\sigma_v$ as ${\cal M} =
\sigma_v/c_{\rm s}$. For each value of the Mach number we consider three
different cases, one case where turbulence is driven on large scales only
(i.e.\ with wavenumbers $k$ in the interval $1\le k \le 2$),
intermediate-wavelength turbulence ($3\le k \le 4$), and small-scale
turbulence ($7 \le k \le 8$), as summarized in Table \ref{tab:models}. Note
that our models are not subject to global shear because of the adopted
periodic boundary conditions. We call turbulence "large scale" when the
Fourier decomposition of the velocity field is dominated by the largest scales
possible for the considered volume $L^3$, i.e.\ the system becomes isotropic
and homogeneous only on scales larger than $L$. On scales below $L$ it may
exhibit a considerable degree of anisotropy. This is most noticeable in the
case $1\le k \le 2$, because wavenumber space is very poorly sampled and
variance effects become significant. The system is dominated by one or two
large shock fronts that cross through the medium. In the interval $7\le k \le
8$ the number of Fourier modes contribution to the velocity field is large,
and the system appears more isotropic and homogeneous already on distances
smaller than $L$. This trend is clearly visible in Figure
\ref{fig:3Dplot-cut}.

{
 \tiny
%
\begin{table}[htp]
{\caption{Model properties}
\label{tab:models}
}
\scriptsize
\begin{center}
\begin{tabular}[t]{ccccccccccccccc}
\hline 
\footnotesize{(1)}  & \footnotesize{(2)}  & \footnotesize{(3)} &
\footnotesize{(4)}  & \footnotesize{(5)}  & \footnotesize{(6)} &
\footnotesize{(7)}  & \footnotesize{(8)}  & \footnotesize{(9)} &
\footnotesize{(10)} & \footnotesize{(11)} & \footnotesize{(12)}&
\footnotesize{(13)} \\
   {model}     & {$k$}     & {${\cal M}$}  & {$t_{\rm cross}$}   & {$\bar{\sigma}_x$} 
 & {$\bar{\sigma}_y$}   & {$\bar{\sigma}_z$}   & $D_x(\infty)$   & $D_y(\infty)$   & $D_z(\infty)$
 & $2\bar{\sigma}_x/k$  & $2\bar{\sigma}_y/k$  & $2\bar{\sigma}_z/k$\\
%
%
\hline 
{\em 0}$\ell$ & $1..2$ &   0.6 &  35.3 & 0.030 & 0.028 & 0.027 & 0.027
& 0.021 & 0.019 & 0.030 -- 0.060 & 0.028 -- 0.057 & 0.027 -- 0.054 \\
{\em 0i} & $3..4$ &   0.5 &  39.1 & 0.026 & 0.026 & 0.025 & 0.010 &
0.010 & 0.009 & 0.013 -- 0.017 & 0.013 -- 0.017 & 0.013 -- 0.017 \\
{\em 0s} & $7..8$ &   0.4 &  46.2 & 0.021 & 0.022 & 0.022 & 0.005 &
0.005 & 0.005 & 0.005 -- 0.006 & 0.005 -- 0.006 & 0.005 -- 0.006 \\
\hline
{\em 1}$\ell$ & $1..2$ &   1.9 &  10.4 & 0.106 & 0.084 & 0.098 & 0.140
& 0.069 & 0.111 & 0.106 -- 0.213 & 0.084 -- 0.167 & 0.098 -- 0.196 \\ 
{\em 1i} & $3..4$ &   1.9 &  10.6 & 0.097 & 0.096 & 0.092 & 0.042 &
0.047 & 0.038 & 0.048 -- 0.065 & 0.048 -- 0.064 & 0.046 -- 0.061 \\     
{\em 1s} & $7..8$ &   1.7 &  11.5 & 0.086 & 0.089 & 0.087 & 0.025 &
0.026 & 0.024 & 0.021 -- 0.024 & 0.022 -- 0.025 & 0.022 -- 0.025 \\
\hline 
{\em 2}$\ell$ & $1..2$ &   3.1 &   6.5 & 0.173 & 0.129 & 0.158 & 0.223
& 0.103 & 0.169 & 0.173 -- 0.346 & 0.129 -- 0.257 & 0.158 -- 0.315 \\
{\em 2i} & $3..4$ &   3.1 &   6.4 & 0.167 & 0.155 & 0.151 & 0.084 &
0.071 & 0.063 & 0.083 -- 0.111 & 0.077 -- 0.103 & 0.075 -- 0.100 \\
{\em 2s} & $7..8$ &   3.2 &   6.3 & 0.154 & 0.163 & 0.157 & 0.044 &
0.054 & 0.047 & 0.038 -- 0.044 & 0.041 -- 0.046 & 0.039 -- 0.045 \\
\hline 
{\em 3}$\ell$ & $1..2$ &   5.2 &   3.8 & 0.301 & 0.252 & 0.227 & 0.314
& 0.245 & 0.169 & 0.301 -- 0.603 & 0.252 -- 0.505 & 0.227 -- 0.454 \\ 
{\em 3i} & $3..4$ &   5.8 &   3.5 & 0.261 & 0.287 & 0.316 & 0.131 &
0.189 & 0.233 & 0.130 -- 0.174 & 0.143 -- 0.191 & 0.158 -- 0.211 \\
{\em 3s} & $7..8$ &   5.8 &   3.4 & 0.297 & 0.288 & 0.289 & 0.106 &
0.092 & 0.091 & 0.074 -- 0.085 & 0.072 -- 0.082 & 0.072 -- 0.083 \\
\hline 
{\em 4}$\ell$ & $1..2$ &   8.2 &   2.4 & 0.467 & 0.318 & 0.444 & 0.693
& 0.241 & 0.558 & 0.467 -- 0.933 & 0.318 -- 0.635 & 0.444 -- 0.887 \\
{\em 4i} & $3..4$ &   9.7 &   2.1 & 0.451 & 0.478 & 0.520 & 0.248 &
0.323 & 0.349 & 0.225 -- 0.301 & 0.239 -- 0.319 & 0.260 -- 0.347 \\ 
{\em 4s} & $7..8$ &  10.4 &   1.9 & 0.532 & 0.513 & 0.519 & 0.194 &
0.167 & 0.170 & 0.133 -- 0.152 & 0.128 -- 0.147 & 0.130 -- 0.148 \\ 
\hline
\hline
\end{tabular}
\end{center}
{%
1.\ column: Model identifier,  with the letters  $\ell$, $i$, and $s$ standing for
 large-scale, intermediate-wavelength, and short-wavelength turbulence, respectively.\\
2.\ column: Driving wavelength intervall.\\
3.\ column: Mean Mach number,  defined  
 as ratio between the time-averaged one-dimensional  velocity dispersion
 $\bar{\sigma}_v = 3^{-1/2} (\bar{\sigma}^{\,2}_x + \bar{\sigma}^{\,2}_y  + \bar{\sigma}^{\,2}_y)^{1/2}$ 
  and the isothermal sound speed $c_{\rm s}$, ${\cal M}=
 \bar{\sigma}_v/c_{\rm s}$. The values for the
 different velocity components $x$, $y$, and $z$ may differ
 considerably, especially for large-wavelength turbulence. Please
 recall from Section \ref{sec:num-meth} that the speed of sound is
 $c_{\rm s} = 0.05$, and thus the sound crossing time $t_{\rm sound} =
 20$.\\
4.\ column: Average shock crossing time through the computational volume.\\
5.\ to 7.\ column: Time averaged  velocity dispersion along the
 three principal axes $x$, $y$, and $z$, e.g.\ for the $x$-component
 $\bar{\sigma}_x^{\,2} = \int_0^t \langle (v_{x.i}(t') - \langle
 v_{x.i}(t')  \rangle_i )^2\rangle_i dt' / t$. \\
8.\ to 10.\ column: Mean-motion corrected diffusion coefficients along
 the  three principal axes computed from Equation \ref{eqn:D(t)-1} for
 time intervals $t\gg\tau$.\\
11.\ to 13.\ column: Predicted values of the mean motion corrected
 diffusion coefficients  $D^\prime_x$, $D^\prime_y$, and $D^\prime_z$
 from  extending mixing length theory into the 
 supersonic regime (Section \ref{sec:mixing}).\\
}
\end{table}
}

Similar to any other numerical calculations, the models discussed here fall
short of describing real gases in comprehensive details as they cannot include
all physical processes that may act on the medium.  In interstellar gas
clouds, transport properties and chemical mixing will not only be determined
by the compressible turbulence alone, but the density and velocity structure
is also influenced by magnetic fields, chemical reactions, and radiation
transfer processes.  Furthermore, all numerical models are resolution limited.
The turbulent inertial range in our large-scale turbulence simulations spans
over about 1.5 decades in wavenumber. This range is considerably less than
what is observed in interstellar gas clouds.  The same limitation holds for
the Reynolds numbers achieved in the models, they fall short of the values in
real gas clouds by several orders of magnitude.  Nevertheless, despite these
obvious shortcomings, the results derived here do characterize global
transport properties in interstellar gas clouds and in other supersonically
turbulent compressible flows.

\begin{figure}[ht]
\vspace{0.5cm}
\includegraphics[width=15cm]{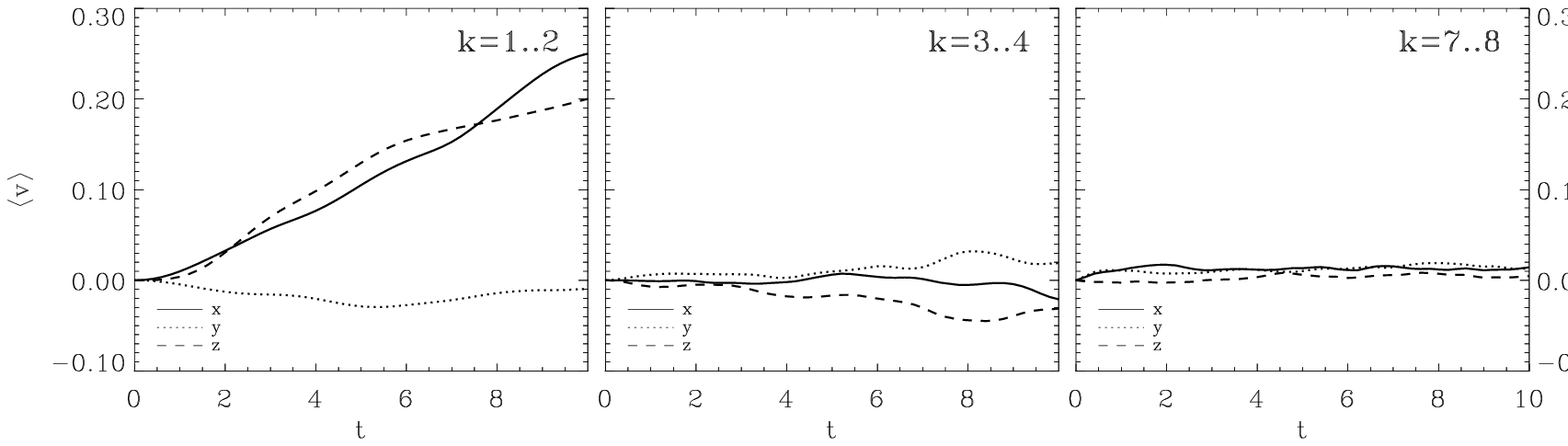}
\vspace{0.5cm}
\caption{\label{fig:v-mean}%
Time evolution of the mean flow velocity $\langle \vec{v}_i(t)
\rangle_i$ in models {\em 2}$\ell$,  {\em 2i},  and {\em 2s}. Time $t$
and velocity $v$ are given in normalized units (Section \ref{sec:num-meth}).}
\end{figure}
\section{Transport properties}
\label{sec:transport}
Supersonic turbulence in compressible media establishes a complex
network of interacting shocks.  Converging shock fronts locally
generate large density enhancements, diverging flows create rarefied
voids of low gas density.  The fluctuations in turbulent velocity
fields are highly transient, as the random flow that creates local
density enhancements can disperse them again. The life time of
individual shock-generated clumps corresponds to the time interval for
two successive shocks to pass through the same location in space,
which in turn depends on the length scale of turbulence and on the
Mach number of the flow.

The velocity field of turbulence that is driven at large wavelengths
is found to be dominated by large-scale shocks  which are very
efficient in sweeping up material, thus creating massive coherent
density structures. The shock passing time is rather long, and shock-generated
clumps can travel quite some distance before begin 
disrupted. 
On the contrary, when energy is inserted mainly on small scales, the network
of interacting shocks is very tightly knit.  Clumps have low masses and the
time interval between two shock fronts passing through the same location is
small, hence, swept-up gas cannot travel far before being
dispersed again. 

The density and velocity structure of three models with large-,
inter\-mediate-, and small-wavelength turbulence is visualized in
Figure~\ref{fig:3Dplot-cut}. It shows cuts through the centers of the
simulated volume. As turbulence is stationary, all times are
equivalent, and the snapshot in the upper panel is taken at some
arbitrary time. The lower panel depicts the system some time interval
later corresponding to $1/4$ shock crossing time through the cube. One
clearly notices markable differences in the density and velocity field
between the three models.

\subsection{Transport properties in an absolute reference frame}
\label{subsec:Euler}
In order to drive supersonic turbulence and to maintain a given rms
Mach number in the flow, we use a random Gaussian velocity field with
zero mean to `agitate' the fluid elements at each timestep. However,
despite the fact that the driving scheme has zero mean, the system is
likely to experience a net acceleration and develop an appreciable
drift velocity, because of the compressibility of the medium. This
evolutionary trend is well illustrated in Figure \ref{fig:v-mean}
which plots the time evolution of the three components of the mean
velocity for models {\em 2}$\ell$, {\em 2i}, and {\em 2s}, with rms
Mach numbers ${\cal M} \approx 3.1$, where turbulence is driven on
{\em (a)} large (i.e.\ with small wavenumbers $1\le k \le2$), {\em
(b)} intermediate ($3\le k \le 4$), and {\em (c)} small scales (with
$7\le k \le 8$). The net acceleration is most pronounced when
turbulent energy is inserted on the global scales, as in this case
larger and more coherent velocity gradients can build up across the
volume compared to small-scale turbulence.

The tendency for the zero-mean Gaussian driving mechanisms to induce
significant center-of-mass drift velocities in highly compressible
media can be understood as follows. Suppose the gas is perturbed by
one single mode in form of a sine wave. If the medium is homogeneous
and incompressible, equal amounts of mass would be accelerated in the
forward as well as in the backward direction. But, if the medium is
inhomogeneous, there would be an imbalance between the two directions
and the result would be a net acceleration of the system.  If the
density distribution remains fixed, this acceleration would be
compensated by an equal amount of deceleration after half a period,
and the center of mass would simply oscillate.  However, if the system
is highly compressible and the driving field is a superposition of
plane waves, the density distribution would change continuously (and
randomly).  Any net acceleration at one instance in time would not be
completely compensated after some finite time interval later. This
will only occur for $t\rightarrow \infty$ assuming ergodicity of the
flow. Subsequently, the system is explected to develop a net flow
velocity in some random direction for $t<\infty$. This effect is most
clearly noticeable for long-wavelength turbulence, where density and
velocity structure is dominated by the coherent large-scale structure.
But the effect is small for turbulence that is excited on small
scales, because in this limit, there is a large number of accelerated
`cells' which in turn compensate for another's acceleration.

\begin{figure}[tp]
\includegraphics[width=15cm]{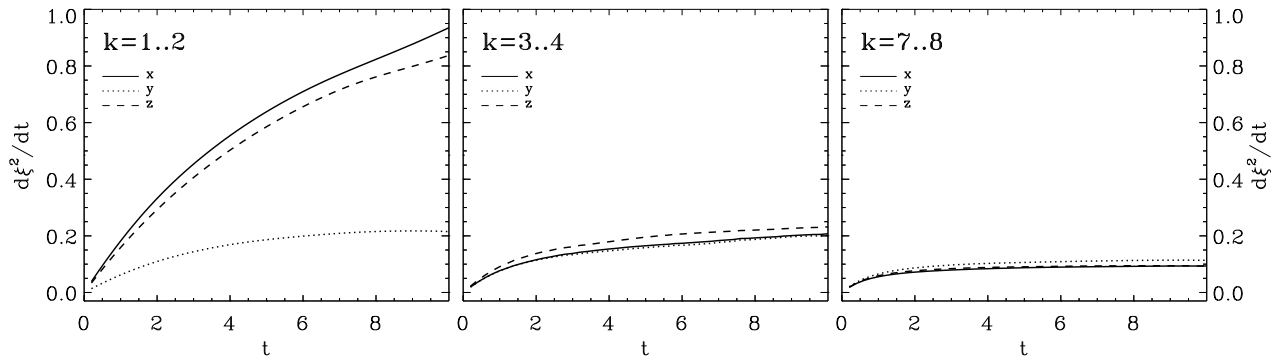}
\vspace{0.5cm}
    \caption{\label{fig:Euler-D}%
      Time evolution of the diffusion coefficient $D(t)=d\langle
      [\vec{r}_i(t)-\vec{r}_i(0)]^2 \rangle_i / dt$ for models {\em
        2}$\ell$, {\em 2i}, and {\em 2s} computed in an absolute
      reference frame. All units are normalized as described in
      Section \ref{sec:num-meth}.
}
\end{figure}
The property that the compressible turbulent flows are likely to
pick up average drift velocities, even when driven by Gaussian fields
with zero mean, has implications for the transport coefficients.
Figure \ref{fig:Euler-D} shows the time evolution of the absolute
(Eulerian) diffusion coefficients $D_x$, $D_y$, and $D_z$ in each
spatial direction computed from Equation \ref{eqn:def-sigma}.
%
Note that for stationary turbulence, only time differences are
relevant and one is free to chose the initial time.  In order to
improve the statistical significance of the analysis, we obtain $D(t)$
and $\xi_{\vec{r}}(t)$ by further averaging over all time intervals
$t$ that `fit into' the full timespan of the simulation.
%
%


Due to the (continuous) net acceleration experienced by the system,
the quantity $\xi_{\vec{r}}^2(t)$ grows faster than linearly with
time, even for intervals much larger than the correlation time $\tau$,
i.e.\ for $\tau \ll t < \infty$.  The system resides in a
superdiffusive regime, where $D(t)$ does not saturate. Instead, $D(t)$
grows continuously with time, which is most evident in model {\em
  2}$\ell$ of large-scale turbulence. The ever increasing drift
velocity $\langle \vec{v}_i(t) \rangle_i$ causes strong velocity
correlations leading continuous growth of the velocity autocorrelation
tensor $\int_0^{t} {\rm tr}\,{\cal C}(t')dt'$.  This net motion,
however, can be corrected for, allowing us to study the dispersion of
particles in a reference frame that moves along with the average flow
velocity of the system.

\begin{figure}[pt]
\vspace{0.5cm}
\includegraphics[width=11.7cm]{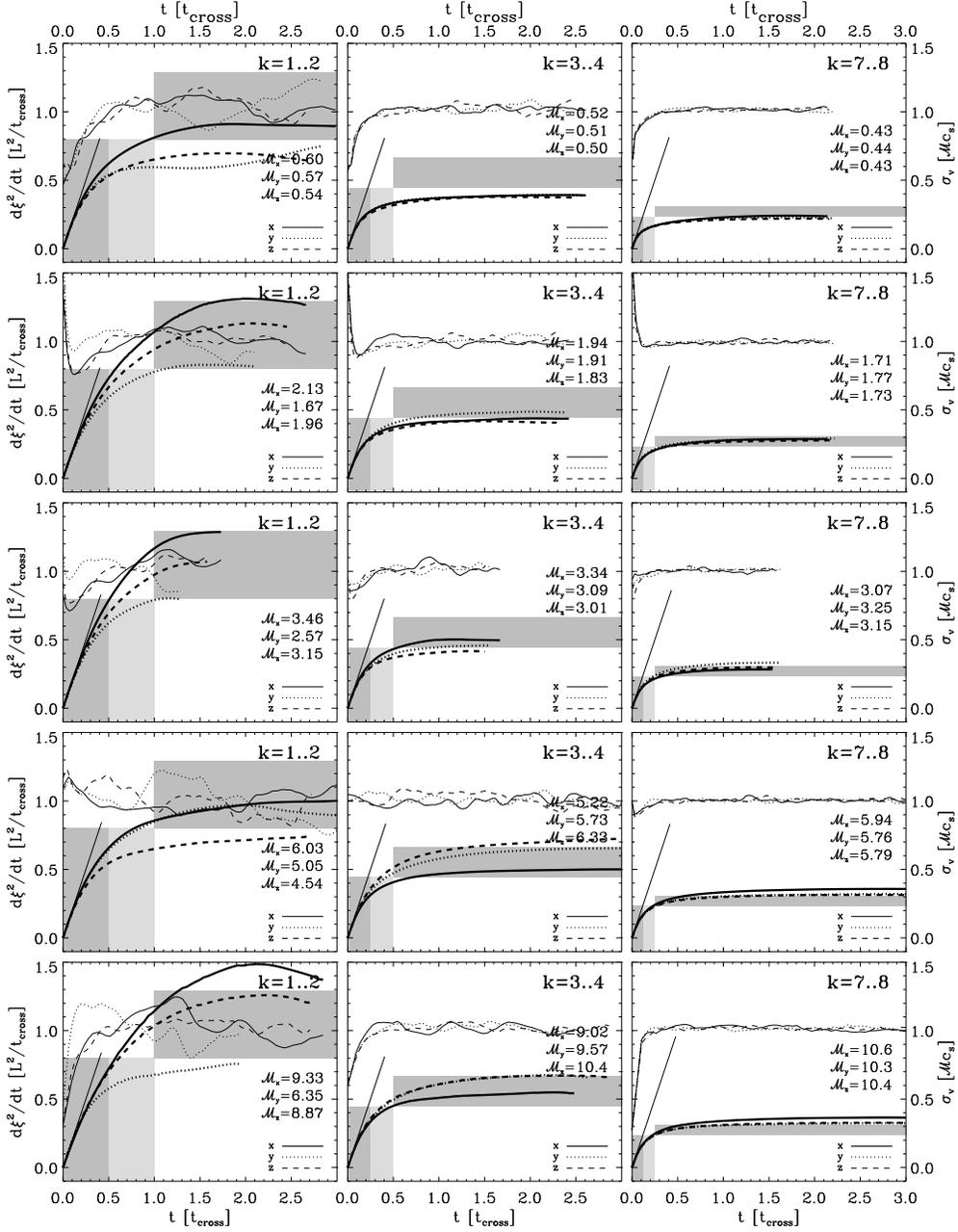}
\vspace{0.5cm}
    \caption{\label{fig:Lagrange-D}%
      Time evolution of the diffusion coefficient $D^\prime(t)=d\xi^2_i / dt$
      computed in a reference frame that follows the average flow velocity
      ({\em thick line}, axis scaling on the left ordinate), i.e.\ is centered
      on $\langle \vec{r}_i(t) \rangle_i = \int_0^t\langle \vec{v}_i(t')
      \rangle_i dt'$.  Velocity dispersions along the three major axes $x$,
      $y$, and $z$ are each normalized to unity using the time-averaged
      one-dimensional Mach number $\cal M$ (as indicated in each plot)
      together with the given value of the sound speed $c_{\rm s}$ ({\em thin
        lines}, axis scaling on the right ordinate). Times are rescaled to
      the rms shock crossing time through the simulated cube $t_{\rm cross} =
      L/\sigma_v = L/({\cal M}c_{\rm s})$. Details for each model are given in
      Table \ref{tab:models}. The horizontal gray shaded area indicates the
      mixing length prediction for $t\rightarrow \infty$, and the vertical
      gray and light gray shaded areas show a time interval of
      $\tau=L/(k{\cal M}c_{\rm s})$ and $2\tau$, respectively. For $t\ll\tau$
      diffusion should be anomalous and coherent, with $D^\prime(t)$ growing
      linearly with time. The expected behavior from mixing length theory in
      this regime is indicated by the straight line originating at $t=0$. It
      indeed gives a good fit. Note that all models driven on large scales
      ($1\le k \le 2$) exhibit a considerable degree of anisotropy, manifested by
      different rms Mach numbers $\cal M$ along the three principal axes and
      different values $D^\prime(t)$. For the models with intermediate-scale
      and small-scale driving anisotropy effects are increasingly less
      important.}
\end{figure}

\subsection{Transport properties in flow coordinates}
\label{subsec:Lagrange}

In order to gain further insight into the transport properties of
compressible supersonic turbulent flows, we study the time evolution
of the relative (Lagrangian) diffusion coefficient. In this
prescription, 
\begin{equation}
D^\prime (t)= \frac{d \xi^2_{\vec{r}}(t)}{dt} \nonumber
\end{equation} 
is obtained relative to a frame of reference which comoves with the
mean motion of the system $\langle \vec{v}_i(t) \rangle_i$ on the
trajectory $\langle \vec{r}_i(t) \rangle_i = \int_0^t\langle
\vec{v}_i(t') \rangle_i dt'$. Then, $\xi^2_{\vec{r}}(t-t')= \langle
[(\vec{r}_i(t)-\langle \vec{r}_i(t) \rangle_i)-(\vec{r}_i(t')- \langle
\vec{r}_i(t') \rangle_i)]^2 \rangle_i$ (see Equation
\ref{eqn:def-sigma}).

In Figure \ref{fig:Lagrange-D}, we show the evolution of $D^\prime
(t)$ for each coordinate direction for the complete suite of models.
The rms Mach numbers range from about 0.5 to 10, each considered for
three cases where turbulence is driven on large, intermediate, and
small scales, respectively. The plots are rescaled such that the
time-averaged one-dimensional rms velocity dispersion $\bar{\sigma}_{v}$ is
normalized to unity (for each direction separately). We also rescale
the time $t$ with respect to the average shock crossing time
scale through the computational volume, $t_{\rm cross} =
L/\bar{\sigma}_{v}$.  Recall that $L=1$, and note that $\bar{\sigma}
_{v}$ usually differs between the three spatial directions because of
the variance effects, especially in models of large-scale turbulence.

In Figure \ref{fig:Lagrange-D}, we demonstrate that the magnitude of
$D^\prime (t)$ saturates for large time intervals $t> \tau$ in all
directions. In a reference frame that follows the mean motion of the
flow, diffusion in compressible supersonically turbulent media indeed
behaves in a normal manner.  For small time intervals $t< \tau$,
however, the system still exhibits an anomalous diffusion even with
the mean-motion correction. In this regime $D^\prime(t)$ grows roughly
linearly with time. For $t>\tau$ the diffusion coefficient $D^\prime
(t)$ reaches an asymptotic limit. This result holds for the entire
range of Mach numbers studied and for turbulence that is maintained by
energy input on very different spatial scales.

From Figure \ref{fig:Lagrange-D}, we find that diffusion in
compressible supersonic turbulent flows follows a universal law.  It
can be obtained by using the rms Mach number (together with the sound
speed $c_{\rm s}$) as characterizing parameter for rescaling the
velocity dispersion $\sigma_v$, and the rms shock crossing time scale
through the volume $t_{\rm cross} = L/({\cal M}c_{\rm s})$ for
rescaling the time.  The normalized diffusion coefficient $D^\prime(t)$
exhibits a universal slope of two at times $t<\tau$ (i.e.\ in the
superdiffusive regime), and approaches a constant value that depends
only on the length scale but not on the strength (i.e.\ the resulting
Mach number) of the mechanism that drives the turbulence.  Even for
highly compressible supersonic turbulent flows it is possible to find
simple scaling relations to characterize the transport properties ---
analogous to the mixing length description of diffusive processes in
incompressible subsonic turbulent flows.

\section{A mixing length description}
\label{sec:mixing}
Incompressible turbulence is often described in terms of a hierarchy
of turbulent eddies, where each eddy contains multiple eddies of
smaller size on the lower levels of the hierarchy, while itself being
part of turbulent eddy at larger scales \cite{R22, K41,O41}. At each
level of the hierarchy, an eddy is characterized by a typical
lengthscale $\tilde{\ell}$ and a typical velocity $\tilde{v}$. The
typical lifetime of an eddy is its `turn-over' time $\tau=
\tilde{\ell} / \tilde{v}$.  This mixing length prescription is an
attempt to characterize the flow properties in terms of the typical
scales $\tilde{\ell}$ and $\tilde{v}$. For example, this classical
picture defines an effective `eddy' viscosity $\mu = \rho
\tilde{\ell}\tilde{v}$, where $\rho$ is the density. The mixing length
$\tilde{\ell}$ is interpreted to be the turbulent analogue of the mean
free path of molecules in the kinetic theory of gases, with
$\tilde{v}$ being the characteristic velocity of the turbulent
fluctuation. 

In such a model, the velocities of gas molecules within an eddy are
strongly correlated within a time interval $t<\tau$. They all follow
the eddy rotation; the diffusion process is coherent.  However, for
$t\gg\tau$ the velocities of gas molecules become uncorrelated, as the
eddy has long been destroyed and dispersed. Hence, the velocity
autocorrelation function vanishes for large time intervals, ${\cal
C}(t)\rightarrow 0$ for $t\rightarrow \infty$. Diffusion becomes
incoherent as in Brownian motion or the random walk.  The diffusion
coefficient in the mixing length approach simply is $D(t) \approx 2
\tilde{v}^2 t$ in the regime $t<\tau$, which follows from replacing
$\vec{r}(t)$ by $\tilde{v}t$ and $\vec{v}(t)$ by $\tilde{v}$ in
Equation \ref{eqn:D(t)-1}. As the largest correlation length is the
eddy size, $\vec{r}(t)$ is substituted by $\tilde{\ell}=\tilde{v}\tau$
for times $t\gg\tau$, and the classical mixing length theory yields
$D(t) \approx 2 \tilde{\ell} \tilde{v} = 2 \tilde{v}^2 \tau = {\rm
constant}$.

Compressible, supersonic, turbulent flows rapidly build up a network
of interacting shocks with highly transient density and velocity
structure. Density fluctuations are generated by locally converging
flows, and their lifetimes are determined by the time $\tau$ between
two successive shock passages. This time interval is determined by the
typical shock velocity, which is roughly the rms velocity of the flow,
i.e.\ the Mach number times the sound speed, $\sigma_v = {\cal M}
c_{\rm s}$. It also depends on the length scale at which energy is
inserted into the system to maintain the turbulence, which in our case
is $L/k$ with $k$ being the driving wavenumber and $L$ being the size
of the considered region (recall that in our models $L$ is
unity). This length scale is also the typical traveling distance
before two shocks interact with each other.  As basic ingredients for
a supersonic compressible mixing length description we can thus
identify:
\begin{eqnarray}
\label{eqn:mixing-length}
{\rm shock~travel~length:}&&\tilde{\ell}  \approx  L/k , \\
{\rm rms~velocity:} &&\tilde{v}  \approx  \sigma_v = {\cal M} c_{\rm s} .
\end{eqnarray}
The Lagrangian  velocity correlation time scale, $\tau$, is
analogeous to the time interval during which shock-generated density
fluctuation remains unperturbed and moves coherently before it is
being dispersed by the interaction with a new shock front. This time
interval is equivalent to the time scale a shock travels along its
`mean free path' $\tilde{\ell}$ with an rms velocity $\tilde{v}$. This
crossing time is $\tau = \tilde{\ell}\tilde{v} \approx \sigma_v\,L/k$.
For $t<\tau$ gas molecules can travel coherently within individual
shock generated density fluctuations, and the diffusion coefficient in
the mixing length prescription follows as
\begin{equation}
  \label{eqn:D(t)-small-t}
  D^\prime(t) \approx 2\tilde{v}^2t \approx 2\sigma_v^2 t\;.
\end{equation}
$D^\prime(t)$ grows linearly with time with slope $2\sigma_v^2$.
For large times, $t\gg\tau$, $D^\prime(t)$ approaches a constant value,
\begin{equation}
  \label{eqn:D(t)-large-t}
  D^\prime (t) \approx 2\tilde{v}^2\tau \approx 2\sigma_v\,L/k\;.
\end{equation}

This mixing length approach (Equations \ref{eqn:D(t)-small-t} and
\ref{eqn:D(t)-large-t}) suggests a unique scaling dependence of the
diffusion coefficients in supersonic compressible flows on the {\em
Mach number} $\cal M$ and on the {\em length scale} $\tilde{\ell}$ of
the most energy containing modes {\em with respect to the total size
$L$ of the system considered}.

We can use $\cal M$ (together with the given value of the sound speed)
to normalize the rms velocity: $\sigma_v= {\cal M} c_{\rm s} \mapsto
\sigma_v'=1$. And we can also rescale the time with respect to the rms
shock crossing time scale through the total volume, which is $t_{\rm
cross} = L/\sigma_v = L/({\cal M} c_{\rm s}) = t_{\rm sound}/{\cal M}$
with $t_{\rm sound} = L/c_{\rm s}$ being the sound crossing time, so that
$t \mapsto t'=t/t_{\rm cross}$. From this normalization procedure, we
get $D^\prime (t)\mapsto D''(t') = D^\prime (t) \,{\cal M} c_{\rm s}L$
and obtain the following universal profile for the diffusion
coefficient,
\begin{eqnarray}
\label{eqn:scaled-D1}
 D^{\prime\prime}(t')=2t' & {\rm ~for~}\; & t' \ll 1/k {\rm ~~~~and}\\
\label{eqn:scaled-D2} D^{\prime\prime}(t')=2/k & {\rm ~for~}\; & t' \gg 1/k \;.  
\end{eqnarray}

Note that this result holds for each velocity component separately, as
the results in Figure \ref{fig:Lagrange-D} indicate. In this case
$\sigma_v$ stands for $\sigma_x$, $\sigma_y$, or $\sigma_z$ in
Equations \ref{eqn:D(t)-small-t} and \ref{eqn:D(t)-large-t}, and it
hold for the total diffusion coefficient, when using $\sigma_v =
({\sigma_x^2 + \sigma_y^2 + \sigma_z^2})^{1/2}$ instead.

\begin{figure}[htp]
\vspace{0.5cm}
\includegraphics[width=15cm]{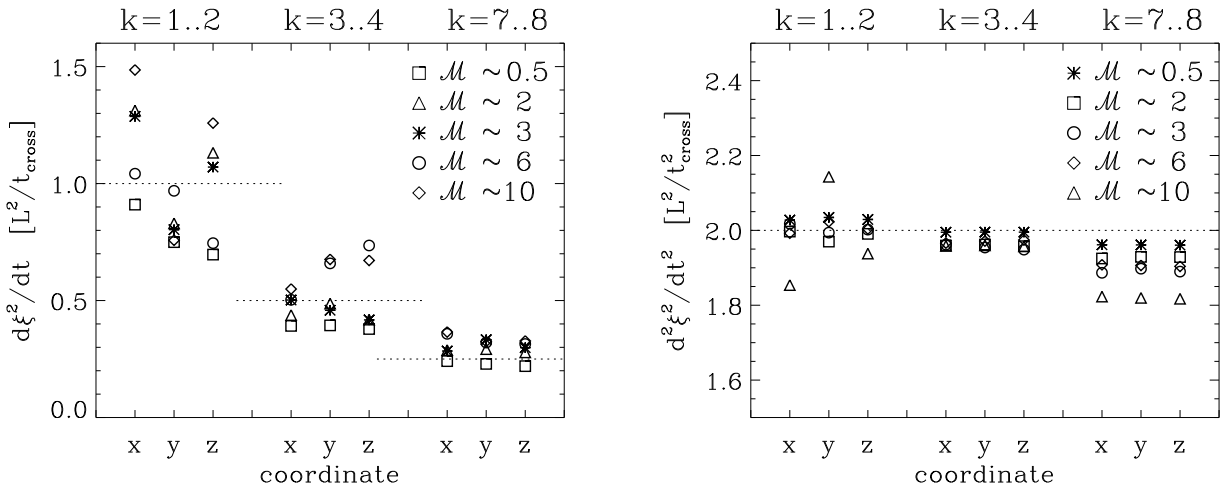}
\vspace{0.9cm}
    \caption{\label{fig:comparison}%
      Comparison between mixing length predictions and numerical
      models.  At the left we plot he normalized, mean-motion
      corrected diffusion coefficient $D''(t')$ for $t\rightarrow
      \infty$, and at the right its slope $dD''(t')/dt'$ for $t'\ll
      1/k$. For each suite of models, large-scale, intermediate-scale,
      and small-scale turbulence, respectively (as indicated by the
      forcing wavenumber $k$ at the top of each plot), we separately
      show the three velocity components (as indicated at the bottom).
      The different Mach numbers in each model suite are denoted by
      different symbols (as identified at the right-hand side of each
      plot). The dotted lines give the corresponding prediction of the
      mixing-length theory, $D''(t')=2/k$ and $dD''(t')/dt'=2$,
      respectively, where we take $k$ to be the maximum wavenumber of the
      forcing scheme (indicated at the top of each plot).}
\end{figure}
The validity of the mixing length approximation is quantified in
Figure \ref{fig:comparison} which plots the mixing length predictions
against the values obtained from the numerical models. For large-scale
and intermediate scale turbulence, the mixing length approach gives
very satisfying results, only for small-scale turbulence it
underestimates the diffusion strength. This disparity probably has to
do with the numerical resolution of the code, in the sense that
driving wavenumbers of $k\approx 8$ come close to the dissipation
scale of the method and hence the inertial range of turbulence is
limited \cite{KHM00}. That limitation leads to an effective driving
for the turbulent motion on somewhat larger scales than $1/8$.
Consequently, it leads to a stronger diffusion, i.e.\ somewhat larger
diffusion coefficients than those predicted by Equations
\ref{eqn:scaled-D1} and \ref{eqn:scaled-D2}. The same numerical effects also account for the
slightly shallower slope of $D'(t)$ for $t\ll\tau$ for models $7\le k\le 8$.

Figures \ref{fig:Lagrange-D} and \ref{fig:comparison} indicate that
the classical mixing length theory can be extended from incompressible
(subsonic) turbulence into the regime of supersonic turbulence of
highly compressible media. In this case, driving length $\tilde{\ell}$
and rms velocity dispersion $\sigma_v= {\cal M}c_{\rm s}$ act as
characteristic length and velocity scales in the mixing length
approach. Note, that this only applies to mean-motion corrected
transport. In general (i.e.\ in an absolute reference frame),
supersonic turbulence in compressible media leads to superdiffusion as
visualized in Figure \ref{fig:Euler-D}.

\section{Summary}
\label{sec:summary}
We studied diffusion processes in supersonically turbulent,
compressible media. To drive turbulence and maintain the desired rms
Mach number in the flow, we insert energy into the system at a
pre-specified rate and over a given spatial scale using random
Gaussian fields. In our numerical models, the adopted magnitude of the
rms Mach numbers range from ${\cal M}=0.5$ to ${\cal M}=10$, and
turbulence was driven on large, intermediate, and small scales,
respectively.

Supersonic turbulence in compressible media establishes a complex
network of interacting shocks.  Converging shock fronts locally
generate  large density enhancements, diverging flows create voids of
low gas density.  The fluctuations in turbulent velocity
fields are highly transient, as the random flow that creates local
density enhancements can disperse them again.

Due to compressibility, supersonically turbulent flows will usually
develop noticeable drift velocities, especially when turbulence is
driven on large scales, even when it is excited with Gaussian fields
with zero mean.  This tendency has consequences for the transport
properties in an absolute reference frame. The flow exhibits
super-diffusive behavior (see also \cite{B01}). However, when the
diffusion process is analyzed in a comoving coordinate system, i.e.\ 
when the induced bulk motion is being corrected, the system exhibits
normal behavior. The diffusion coefficient $D(t)$ saturates for large
time intervals, $t\rightarrow \infty$.

By extending classical mixing length theory into the supersonic regime
we propose a simple description for the diffusion coefficient based on
the rms velocity $\tilde{v}$ of the flow and the typical shock travel
distance $\tilde{\ell}$,
\begin{eqnarray}
D^{\prime}(t) = 2 \tilde{v}^2t & {\mbox{for}} & t\ll
\tilde{\ell}/\tilde{v} \nonumber\,,\\ 
D^{\prime} (t) = 2
\tilde{v}\tilde{\ell} & {\mbox{for}} & t\gg \tilde{\ell}/\tilde{v}
\nonumber\,.
\end{eqnarray}

This functional form may be used in those numerical models where knowledge of
the mixing properties of turbulent supersonic flows is required, but where
these flows cannot be adequately resolved. This is the case, for example, in
astrophysical simulations of galaxy formation and evolution, where the
chemical enrichment of the interstellar gas and the distribution and spreading
of heavy elements produced from massive stars throughout galactic disks needs
to be treated without being able to follow the turbulent motion of
interstellar gas on small enough scales relevant to star formation \cite{R91,
  B98}. Our results furthermore are directly relevant for understanding the
properties of individual star-forming interstellar gas clouds within the disk
of our Milky Way. These are dominated by supersonic turbulent motions which
can provide support against gravitational collapse on global scales, while at
the same time produce localized density enhancements that allow for collapse,
and thus stellar birth, on small scales. The efficiency and timescale of star
formation in galactic gas clouds depend on the intricate interplay between
their internal gravitational attraction and their turbulent energy
content\cite{MK03}.  The same is true for the statistical properties of the
resulting star clusters. For example, the element abundances in young stellar
clusters are found to be very homogeneous \cite{W02}, implying that the gas
out of which these stars formed must be have been chemically well mixed
initially.  On the basis of the results discussed here, this observation can
be used to constrain astrophysical models of interstellar turbulence in
star-forming regions. Understanding transport processes and element mixing in
supersonic turbulent flows thus is a prerequisite for gaining deeper insight
into the star formation phenomenon in our Galaxy.


\acknowledgements{We thank Javier Ballesteros-Paredes, Peter Bodenheimer,
  Mordecai-Mark Mac~Low, and Enrique V{\'a}zquez-Semadeni 
  for many stimulating discussions.  RSK acknowledges support by the
  Emmy Noether Program of the Deutsche Forschungsgemeinschaft (DFG,
  KL1358/1) and subsidies through a NASA astrophysics theory program
  at the joint Center for Star Formation Studies at NASA-Ames Research
  Center, UC Berkeley, and UC Santa Cruz.}

\newpage

\end{document}